\documentclass[aps,prd]{revtex4-1} 
\usepackage{graphicx}
\usepackage{amsmath}
\usepackage{amssymb}
\usepackage{float}
\usepackage{sidecap}
\usepackage{multirow}
\usepackage[caption = false]{subfig}
\usepackage{xcolor}
\usepackage{tikz}
\usetikzlibrary{fit,calc,positioning,decorations.pathreplacing,matrix}
\usepackage{mwe}
\usepackage{sidecap}

\usepackage{braket}
\usepackage{tabularx,booktabs}
\newcolumntype{Y}{>{\centering\arraybackslash}X}
\usepackage[margin=1in]{geometry} 

\begin{document}
\title{\bf Predictions for the beauty meson spectrum}
\author{Mohammad H. Alhakami}
\affiliation{Department of Physics and Astronomy, College of Science, King Saud University, P. O. Box 2455, Riyadh 11451, Saudi Arabia;\\
Nuclear Science Research Institute, KACST, P.O. Box 6086, Riyadh 11442, Saudi Arabia;\\%
and School of Physics and Astronomy, University of Glasgow, Glasgow, G12 8QQ, United Kingdom}
\date{\today}
\begin{abstract}
We predict the spectrum of the four $1S$ and eight $1P$ nonstrange and strange states in the beauty meson family in the context of effective field theory. By the union of heavy quark effective theory and chiral perturbation theory, the mass formalisms for the heavy-light mesons are defined. Our analysis uses mass expressions involving, for the first time, the full leading self-energy corrections and leading power corrections to the heavy quark and chiral limits. The counterterms present in these expressions are fitted using charm data and then used to predict the masses of the analog beauty meson states. The observed spectrum of the ground state, $B^{(*)}_{(s)}$, and excited, $B_{(s)1}$ and $B^*_{(s)2}$, beauty mesons are well reproduced in our theoretical calculations. The excited scalar, $B^*_{(s)0}$, and axial-vector, $B^\prime_{(s)1}$, beauty mesons have not yet been discovered. Hopefully our predictions may provide valuable clues to further experimental exploration of these missing resonances.

\end{abstract}
\maketitle
\section {Introduction}
The physics of heavy-light mesons is well described by heavy quark symmetry. In the heavy quark (HQ) limit, $m_Q\gg \Lambda_{\text{QCD}}$, the spin of the heavy quark, $s_Q$, decouples from the spin of the light degrees of freedom (light antiquarks and gluons), $s_l$, and both separately become conserved in the strong interaction processes. The light degrees of freedom, in this limit, become blind of heavy quark spin and flavor; accordingly charmed and beauty mesons, as heavy-light meson systems, become degenerate. Heavy-light mesons can be organized in doublets of two states with total angular momentum $J_\pm=s_l\pm s_Q$ and parity $P=(-1)^{l+1}$, where  $s_l=l\pm \frac{1}{2}$ and $l$ is the orbital angular momentum of the light degrees of freedom. Here, our focus is on the heavy meson doublets corresponding to $l=0,1$. For the ground state, $l = 0$ ($S$ wave in the quark model), the heavy mesons with $J^P=0^-,1^-$ are degenerate and form members of the ground state $\frac{1}{2}^-$ doublet. For the low-lying excited states, $l=1$ ($P$ wave in the quark model), there are two cases for $s_l$; it could be $\frac{1}{2}$ or $\frac{3}{2}$. For the $\frac{1}{2}^+$ doublet, the degenerate states are $0^+$ and $1^+$. The other $P$-wave states, which form members of the $\frac{3}{2}^+$ doublet, are $1^+$ and $2^+$. Although $1^+$ states of $\frac{1}{2}^+$ and $\frac{3}{2}^+$ doublets can mix, they can be distinguished by their strong decays. In the strict HQ limit, the state $1^+$ of the $\frac{3}{2}^+$ doublet can only decay to ground state by $D$-wave pion emission, so it can be discriminated from $1^+$ of the $\frac{1}{2}^+$ doublet, which decays by $S$ wave.

\begin{figure}[h!]
\includegraphics[width = 0.9\textwidth]{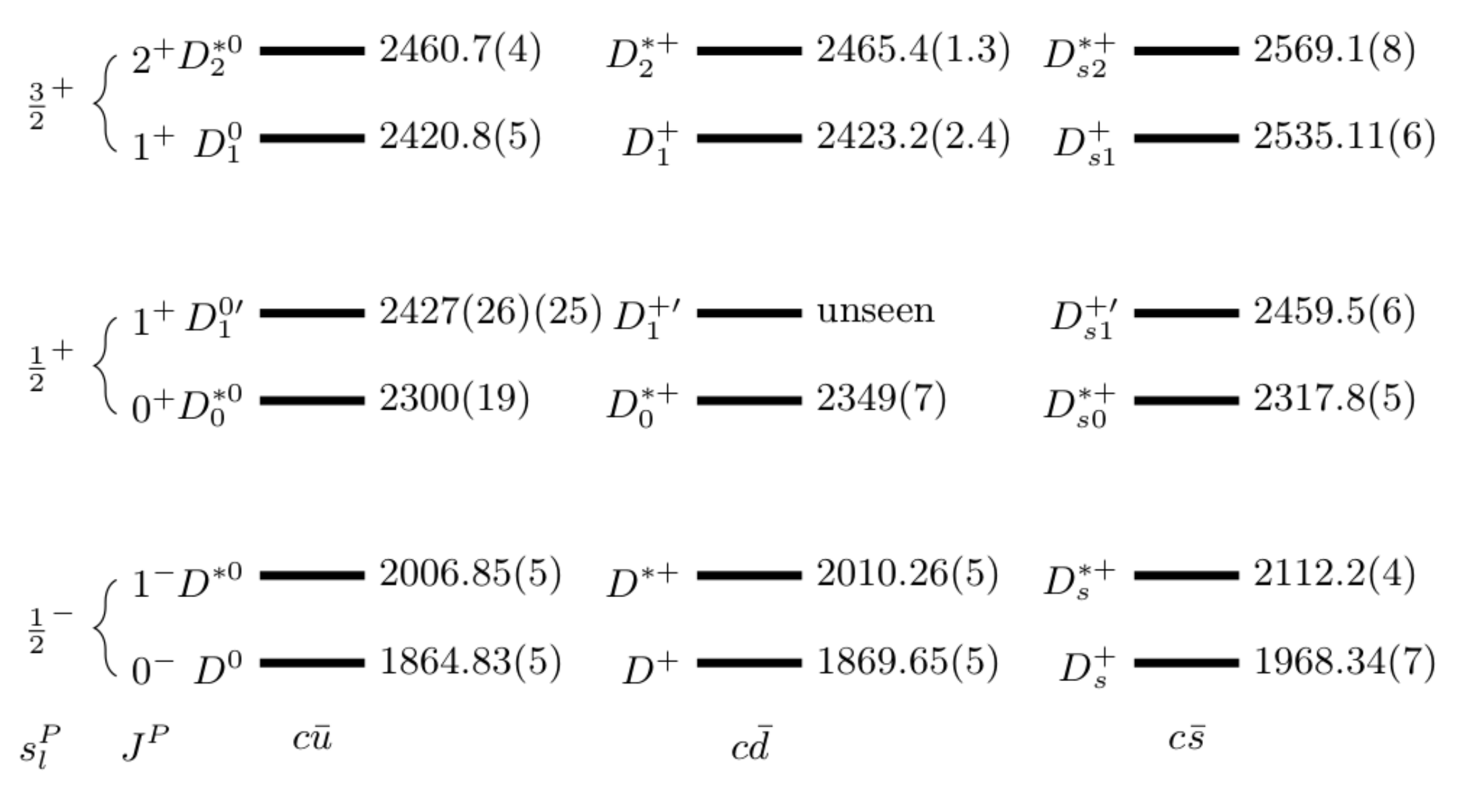}
\caption{The spectrum of charmed mesons. The angular momentum and parity of the meson
(light degrees of freedom) is given by $J^P$ ($s^P_l$).
All masses are taken from the Particle Data Group \cite{pdg12} except the mass of $D^{0\prime}_1$, which is reported by the Belle collaboration \cite{19}.
The masses are given in MeV units. The asterisk is used to refer to states with natural parity $J^P=0^+, 1^-, 2^+$. }
\label{Ddoublets}
\end{figure}

\begin{figure}[h!]
\includegraphics[width = 0.9\textwidth]{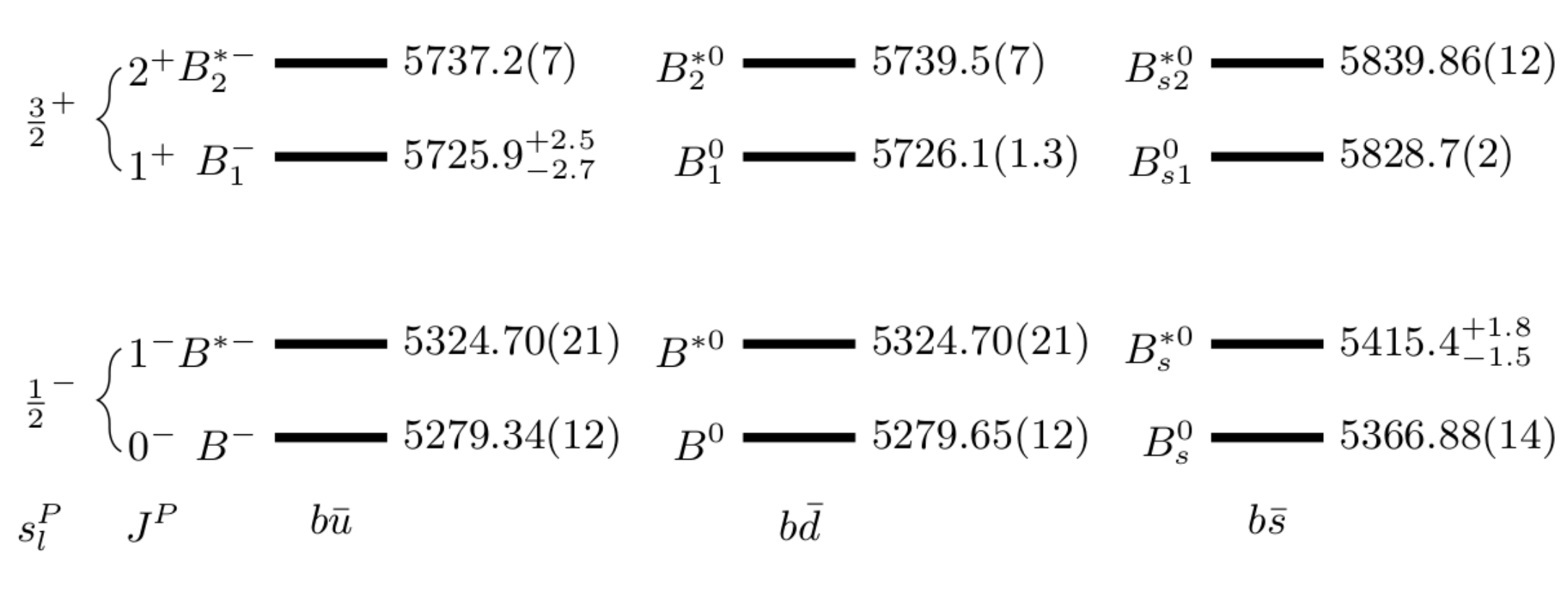}
\caption{The spectrum of beauty mesons. The 
members of the excited $s^P_l=\frac{1}{2}^+$ doublet have not yet been observed. All masses are taken from the Particle Data Group \cite{pdg12}.
For the nonstrange vector meson, we use the same mass for both $B^{*0}$ and 
$B^{*-}$, as given in \cite{pdg12}. The notation is the same as in Fig.~\ref{Ddoublets}}
\label{Bdoublets}
\end{figure}

The measured masses of charmed and beauty mesons are provided in the diagrams given in Figs.~\ref{Ddoublets} and \ref{Bdoublets}. The degeneracy between charm and beauty systems, which is realized in the HQ limit, is in fact lifted by the finiteness of charm and beauty quark masses.
As $m_b>m_c$, the kinetic energy of the heavy quark in the beauty system
is much reduced compared to charm one. This, in turn, significantly reduces the splittings between different doublets in beauty meson system more than their corresponding splittings in charm system;
e.g., $m_{D^{*+}_{s2}}-m_{D^{+}_{s}}=600.76(80)$~MeV whereas $m_{B^{*}_{s2}}-m_{B_{s}}=472.97(21)$~MeV, which indicates the breaking of heavy quark flavor symmetry.   
Additionally, the members of each doublet, i.e.,  $\frac{1}{2}^-$, $\frac{1}{2}^+$, and $\frac{3}{2}^+$, in both charm and beauty sectors, are no longer degenerate, which implies the breaking of heavy quark spin symmetry. The size of such mass splitting, which is called hyperfine splitting, is of order $\Lambda^2_{\text{QCD}}/m_Q$, where $m_Q$ is mass of heavy quark. Consequently, one can relate hyperfine splittings in charm and beauty sectors by a universal factor, which is the ratio of heavy quark masses. 

All the $1S$ and $1P$ charmed mesons are well established, 
as shown in Fig. \ref{Ddoublets}, and hence completing one $S$-wave doublet, $s^P_l=\frac{1}{2}^-$, and two $P$-wave doublets, $s^P_l=\frac{1}{2}^+$ and $s^P_l=\frac{3}{2}^+$.  By examining splitting patterns within charmed meson states, one finds that the mass splittings between strange and nonstrange charmed mesons in the $s_l^P=\frac{1}{2}^-$ and $s_l^P=\frac{3}{2}^+$ doublets are compatible with theoretical expectations of the $SU(3)$ symmetry breaking, which is of order $O(100~\text{MeV})$. 
However, the corresponding splittings in the $s_l^P=\frac{1}{2}^+$ doublet are smaller than the $SU(3)$ violation size.
In fact, the $s_l^P=\frac{1}{2}^+$ charmed strange mesons are well below the threshold for $S$-wave kaon decays to the ground state, 
which deviate very much from quark models predictions; see \cite{rev} and references therein for more detail. These states decay via isospin-violating $\pi^0$ emission, which makes them quite narrow ($\Gamma<3.8$ MeV) unlike their nonstrange counterparts, 
which decay to the ground states by $S$-wave pion emission and accordingly are quite broad ($\Gamma\sim 300$ MeV). Different from the charm sector, only the $1S$ beauty states have been well established, completing one $S$-wave doublet, $s_l^P=\frac{1}{2}^-$. For the $1P$ beauty family, the states belonging to  
$s_l^P=\frac{3}{2}^+$ doublet are well established.
However, the other excited beauty mesons, which belong to $s_l^P=\frac{1}{2}^+$ doublet, have not yet been observed. The current paper is concerned to make model independent predictions for the $1S$ and $1P$ beauty meson spectrum using effective QFT. 

The approximate chiral and heavy quark symmetries of quantum chromodynamics (QCD)
enable us to study the low-energy dynamics of heavy-light meson system. 
The formal approach to employ these symmetries is obtained by constructing effective field theories (EFTs). The heavy quark symmetry is used to build heavy quark effective theory (HQET). This effective theory is potentially a very useful tool in studying masses and 
semileptonic decays of mesons containing a single heavy quark. 
HQET alone, however, is insufficient in studying strong decays of heavy mesons, which involve emission of soft light Goldstone particles.   
For this, HQET and  chiral perturbation theory (ChPT), where the latter describes the low-energy dynamics of the Goldstone particles, are combined in a single framework by introducing heavy meson chiral perturbation theory (HMChPT), for a review see Ref. \cite{cas97}. HMChPT is potentially a very useful tool in analyzing the properties of mesons containing a single heavy quark. We have used this theory to predict the spectrum of the $s^P_l=\frac{1}{2}^-$ and $s^P_l=\frac{1}{2}^+$ beauty mesons \cite{Alhakami}. Our analysis uses the expressions for masses presented in \cite{ms05}.
The leading one-loop corrections between states belonging to the  $s^P_l=\frac{1}{2}^-$ and $s^P_l=\frac{1}{2}^+$ doublets and corrections due to chiral and heavy quark symmetry breakings are considered in our study in \cite{Alhakami}. However, the virtual loops effect from the $s_l^P=\frac{3}{2}^+$ states to $s_l^P=\frac{1}{2}^+$ states are neglected. According to the power counting rules introduced in \cite{ms05}, these missing virtual loop effects are important to the physics of the scalar and axial-vector charmed and beauty mesons. In our recent study \cite{Alhakami20}, we have used a version of HMChPT that includes all relevant heavy quark $s^P_l=\frac{1}{2}^-$, $s^P_l=\frac{1}{2}^+$, and $s^P_l=\frac{3}{2}^+$ doublets. We have calculated the missing loop effects to
the $s^P_l=\frac{1}{2}^+$ masses and also have derived the mass formalisms for the excited $s^P_l=\frac{3}{2}^+$ states, including full one-loop corrections and corrections due to chiral and heavy quark symmetry breaking terms. 

Our main motivation is to extend the applications of HMChPT in \cite{ms05,Alhakami,Alhakami20} to predict the spectrum of the $s^P_l=\frac{1}{2}^-$, $s^P_l=\frac{1}{2}^+$, and $s^P_l=\frac{3}{2}^+$ beauty mesons. Our approach takes into account, for the first time, the full leading self-energy contributions and corrections due to chiral and heavy quark symmetry breakings. 
The mass expressions in \cite{ms05,Alhakami,Alhakami20}
can be used to predict hyperfine and $SU(3)$ flavor mass splittings in different doublets. 
To obtain, however, accurate predictions on the excited 
beauty meson masses, other leading power corrections
to HQ limit of the form $O(\Lambda_{\text{QCD}}/m_c-\Lambda_{\text{QCD}}/m_b)$ are needed.  
In this work, these missing corrections, which give rise the correct mass splittings among the different doublets in the predicted beauty meson spectrum, are properly included to the beauty meson masses. 
The paper is organized as follows. Section II presents the HMChPT masses that we use for the ground state and excited charmed and beauty mesons. The method for fixing unknown parameters using charm spectrum and that for predicting the analog beauty mesons are illustrated in Sec. III. The discussion of our results is also given in this section.
Section IV provides our conclusions.

\section{mass formalisms} 
\begin{SCfigure}
\centering
\caption{A representation for the coupling constants entering the one-loop corrections.
The $g$, $g^\prime$, and $g^{\prime\prime}$ couplings govern the strong interactions among states in the $s^P_l=\frac{1}{2}^-$, $s^P_l=\frac{1}{2}^+$,
and $s^P_l=\frac{3}{2}^+$ doublets, respectively. The $h$ ($h^\prime$) coupling parametrizes the strong interactions of $s^P_l=\frac{1}{2}^-$ and $s^P_l=\frac{1}{2}^+$ ($s^P_l=\frac{1}{2}^+$ and $s^P_l=\frac{3}{2}^+$) mesons.}
\label{coupling}
\includegraphics[width=5cm,height=7cm]{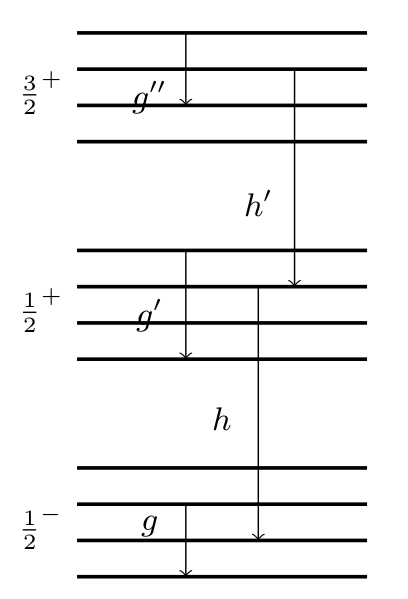}
\end{SCfigure}

The residual masses for the $s^P_l=\frac{1}{2}^-$ ($H$-sector), $s^P_l=\frac{1}{2}^+$ ($S$-sector), and $s^P_l=\frac{3}{2}^+$ ($T$-sector) charmed  mesons have been derived within HMChPT framework including one loop chiral corrections  \cite{ms05,Alhakami20}. They can be written as \cite{Alhakami,Alhakami20}
\begin{equation}
\begin{split}\label{rmassesN}
&m^r_{D_{(s)}}=\eta_H-\frac{3}{4}\xi_H+\alpha_{(s)} L_{H}-\beta_{(s)} F_{H}+\Sigma_{D_{(s)}},\\[1ex]
&m^r_{D^*_{(s)}}=\eta_H+\frac{1}{4}\xi_H+\alpha_{(s)} L_{H}+\frac{1}{3} \beta_{(s)} F_{H}+\Sigma_{D^*_{(s)}},\\[1ex]
&m^r_{D^*_{(s)0}}=\eta_S-\frac{3}{4}\xi_S+\alpha_{(s)} L_{S}-\beta_{(s)} F_{S}+\Sigma_{D^*_{(s)0}},\\[1ex]
&m^r_{D^\prime_{(s)1}}=\eta_S+\frac{1}{4}\xi_S+\alpha_{(s)} L_{S}+\frac{1}{3} \beta_{(s)} F_{S}+\Sigma_{D^\prime_{(s)1}},\\[1ex]
&m^r_{D_{(s)1}}=\eta_T-\frac{5}{8}\xi_T+\alpha_{(s)} L_{T}-\frac{5}{6} \beta_{(s)} F_{T}+\Sigma_{D_{(s)1}},\\[1ex]\
&m^r_{D^*_{(s)2}}=\eta_T+\frac{3}{8}\xi_T+\alpha_{(s)} L_{T}+\frac{1}{2}\beta_{(s)}  F_{T}+\Sigma_{D^*_{(s)2}},
\end{split}
\end{equation}
where $\alpha_{(s)}$ and $\beta_{(s)}$ are $\alpha=-1/3$, $\alpha_s=2/3$, $\beta=-1/4$, $\beta_s=1/2$; the subscript $s$ stands for the strange charmed meson. The $\eta$ and $L$ ($\xi$ and $F$) parameters in $H$, $S$, and $T$ sectors respect (violate) heavy quark spin-flavor symmetry. Note that the parameter $F_X$ (in this work and in \cite{Alhakami20}) has the same definition of the parameter $T_X$ in \cite{Alhakami}; we use $F$ instead of $T$ to avoid confusion with $T$-sector states. Masses in Eq.~\eqref{rmassesN} do not contain terms that only break heavy quark flavor symmetry. The self-energy corrections, which are represented by $\Sigma_{D^\pm}$, are nonlinear functions of the mass difference of charmed mesons and masses of the light pseudoscalar mesons $\pi$, $\eta$, and $K$. They depend quadratically on five ($g$, $g^\prime$,$g^{\prime\prime}$, $h$, and $h^\prime$) couplings. The $g$, $g^\prime$, and $g^{\prime\prime}$ couplings govern the strong interactions among states in the $s^P_l=\frac{1}{2}^-$, $s^P_l=\frac{1}{2}^+$, and $s^P_l=\frac{3}{2}^+$ doublets, respectively. The $h$ ($h^\prime$) coupling parametrizes the strong interactions of $s^P_l=\frac{1}{2}^-$ and $s^P_l=\frac{1}{2}^+$ ($s^P_l=\frac{1}{2}^+$ and $s^P_l=\frac{3}{2}^+$) mesons; for an illustration see Fig.~\ref{coupling}. The explicit expressions of self energies for the excited $s^P_l=\frac{1}{2}^+$ and $s^P_l=\frac{3}{2}^+$ charmed meson states are given in the Appendix of \cite{Alhakami20}. For the $s^P_l=\frac{1}{2}^-$ ground state, we use expressions given in Appendix of \cite{Alhakami}. 
  
The chiral loop functions describing the interaction of heavy mesons with same (opposite) parity are denoted by $K_1$ ($K_2$). They are given, in the $\mathrm{\overline{MS}}$-scheme,  by \cite{Alhakami,Alhakami20}
\begin{equation}\label{K1}
\begin{split}
K_1(\omega, m_i,\mu)&=\frac{1}{16 \pi^2}\left[(-2\omega^3+3m_i^2\omega)\mathrm{ln}\left(\frac{m_i^2}{\mu^2}\right)-4(\omega^2- m_i^2)F(\omega,m_i)+\frac{16}{3}\omega^3-7\omega\, m_i^2\right],\\[2ex]
K_2(\omega, m_i,\mu)&=\frac{1}{16 \pi^2} \left[ (-2\omega^3+ m_i^2\omega)\mathrm{ln}\left(\frac{m_i^2}{\mu^2}\right)-4\omega^2 F(\omega,m_i)+4\omega^3- \omega\, m_i^2\right],
\end{split}
\end{equation}
where the arguments $\omega$, $m_i$, and $\mu$ represent the heavy meson mass differences, the masses of the Goldstone bosons, and the renormalization scale, respectively. The function $F(\omega,m_i)$ is given by
\begin{equation}
 F(\omega,m_i)=\left\{ \begin{array}{c c}-\sqrt{m_i^2-\omega^2} \cos^{-1}(\frac{\omega}{m_i}), & \mbox{$m_i^2>\omega^2,$} \\[2ex]
                                 \sqrt{\omega^2-m_i^2}[i \pi-\cosh^{-1}(-\frac{\omega}{m_i})], & \mbox{$\omega<-m_i,$}\\[2ex]
 \sqrt{\omega^2-m_i^2}\cosh^{-1}(\frac{\omega}{m_i}), & \mbox{$\omega>m_i.$}
                                \end{array} \right.
                                  \end{equation}

The HMChPT results for charmed mesons, Eq.~\eqref{rmassesN}, can be used to obtain the predictions for the analog beauty meson spectrum. For this, the heavy quark spin violating ($\xi$ and $F$) parameters should be rescaled by  $\frac{m_c}{m_b}$. Following Ref. \cite{Alhakami}, we use $\mathrm{\overline{MS}}$ masses, $m_b=4.18$ GeV and $m_c=1.27$ GeV, to define the rescaling factor $m_c/m_b=0.304(50)$, where an extra uncertainty of $O(\Lambda_{\text{QCD}})$ is added to cover the spread of $b$ and $c$ masses resulting from different schemes \cite{pdg12}. The HMChPT masses for the beauty mesons read
\begin{equation}
\begin{split}\label{rmassesNb}
&m^r_{B_{(s)}}=\eta_H-\frac{3}{4}\xi^b_H+\alpha_{(s)} L_{H}-\beta_{(s)} F^b_{H}+\Sigma_{B_{(s)}},\\[1ex]
&m^r_{B^*_{(s)}}=\eta_H+\frac{1}{4}\xi^b_H+\alpha_{(s)} L_{H}+\frac{1}{3} \beta_{(s)} F^b_{H}+\Sigma_{B^*_{(s)}},\\[1ex]
&m^r_{B^*_{(s)0}}=\eta_S-\frac{3}{4}\xi^b_S+\alpha_{(s)} L_{S}-\beta_{(s)} F^b_{S}+\Sigma_{B^*_{(s)0}},\\[1ex]
&m^r_{B^\prime_{(s)1}}=\eta_S+\frac{1}{4}\xi^b_S+\alpha_{(s)} L_{S}+\frac{1}{3} \beta_{(s)} F^b_{S}+\Sigma_{B^\prime_{(s)1}},\\[1ex]
&m^r_{B_{(s)1}}=\eta_T-\frac{5}{8}\xi^b_T+\alpha_{(s)} L_{T}-\frac{5}{6} \beta_{(s)} F^b_{T}+\Sigma_{B_{(s)1}},\\[1ex]\
&m^r_{B^*_{(s)2}}=\eta_T+\frac{3}{8}\xi^b_T+\alpha_{(s)} L_{T}+\frac{1}{2}\beta_{(s)}  F^b_{T}+\Sigma_{B^*_{(s)2}},
\end{split}
\end{equation} 
where $\xi^b_X=\frac{m_c}{m_b}\xi_X$ and $F^b_X=\frac{m_c}{m_b}F_X$. The one-loop corrections, $\Sigma_{B^\pm}$, are now nonlinear functions of the beauty meson mass differences and masses of the light Goldstone particles.

By fitting parameters in Eq.~\eqref{rmassesN} to charmed spectrum, we can use Eq.~\eqref{rmassesNb} to predict the mass splittings in the  beauty sector. To extract the absolute masses of the excited beauty mesons, other leading power corrections $O(\Lambda_{\text{QCD}}/m_c-\Lambda_{\text{QCD}}/m_b)$ to the HQ limit should be included to HMChPT masses. Such missing terms are needed to get the correct mass splittings among the different doublets in the predicted beauty meson spectrum. 
We will elaborate how to properly include them to our masses. For this, let us first recall heavy meson masses in heavy quark effective theory (HQET). In a compact form, the mass of a heavy meson $X$ containing a single heavy quark flavor $Q$ can be expressed up to the leading power corrections to the HQ limit as \cite{hqet} 
\begin{equation}\label{hqet}
m_{X^{(Q)}_\pm}=m_Q+\bar{\Lambda}^X-\frac{\lambda_{X,1}}{2m_Q}\pm n_\mp\frac{\lambda_{X,2}}{2m_Q}, 
\end{equation}
where $n_\pm=2J_\pm+1$ gives the number of spin states in the meson
$X_\pm$. The energy of the light degrees of freedom in the HQ limit is represented by the nonperturbative parameter $\bar{\Lambda}^X$. In the $SU(3)$ limit, this parameter has the same value for all particles in a given $s^P_l$ doublet. If $SU(3)$ breaking is considered, it is different for nonstrange, $\bar{\Lambda}_n^X$, and strange, $\bar{\Lambda}_s^X$, mesons. The other nonperturbative parameters $\lambda_{X,1}$ and $\lambda_{X,2}$ determine the heavy quark kinetic energy and the chromomagnetic energy, respectively.
They have the same values for all particles in a given $s^P_l$ heavy quark doublet. As kinetic energy is a positive quantity, the sign of $\lambda_{X,1}$ in Eq.~\eqref{hqet} should be negative. 

In Eq.~\eqref{hqet}, the third term, which contains $\lambda_{X,1}$, breaks heavy quark flavor symmetry, but it leaves the heavy quark spin symmetry intact. However, the last term, which has $\lambda_{X,2}$, breaks both heavy quark flavor and spin symmetries. The spin averaged mass, $\bar{m}_X$, weighted by the number of helicity states
\begin{equation}\label{avg}
\bar{m}^{(Q)}_X=\frac{n_-m_{X^{(Q)}_-}+n_+m_{X^{(Q)}_+}}{n_++n_-}, 
\end{equation}
is independent of spin symmetry violating parameters, i.e., $\lambda_{X,2}$ ($\xi$ and $F$) in Eq.~\eqref{hqet}
[Eq.~\eqref{rmassesN}]. By using Eqs.~\eqref{hqet} and \eqref{avg}, one can define the difference of spin averaged masses in the beauty sector,
\begin{equation}\label{dbavg}
\begin{split}
\bar{m}^{(b)}_{A_{(s)}}-\bar{m}^{(b)}_{H_{(s)}}&=\bar{m}^{(c)}_{A_{(s)}}-\bar{m}^{(c)}_{H_{(s)}}+\delta^{(s)}_{AH},
\end{split}
\end{equation}
where $A\in \{S,T\}$, and 
\begin{equation}\label{dbavgg}
\delta_{AH}^{(s)}=(\lambda_{A,1}-\lambda_{H,1})^{(s)}\left(\frac{1}{2m_c}-\frac{1}{2m_b}\right), 
\end{equation}
represents the leading $O(\Lambda_{\text{QCD}}/m_c-\Lambda_{\text{QCD}}/m_b)$ corrections. Such corrections to the HQ limit are missing in HMChPT formalisms [Eq.~\eqref{rmassesN}]. Therefore, one has to incorporate them  properly into HMChPT masses for the beauty mesons. It should be noted that the HMChPT masses in Eq.~\eqref{rmassesN} not only involve  effects to first order in the inverse heavy quark mass, $1/m_Q$, as those of HQET in Eq.~\eqref{hqet},  but also involve effects due to the light quark mass, $m_q$, and $m_q/m_Q$ terms. These terms, which are buried in the ($\eta$, $\xi$, $L$, $F$) parameters of Eq.~\eqref{rmassesN}, scale as $\Lambda_{\text{QCD}}^2/m_Q\propto \Delta \sim Q$ and $m_q\sim Q^2$,  where $Q\sim m_\pi,~m_K, m_\eta$ and $\Delta$ is hyperfine splitting operator; for technical details see Refs. \cite{Alhakami,Alhakami20}. As HMChPT masses are defined up to third order, we add the factor,
\begin{equation}\label{dbavggg}
\begin{split}
\Delta^{(s)}_{AH}=\delta^{(s)}_{AH}\left(1+a\frac{M}{\Lambda_\chi}+b\frac{M^2}{\Lambda^2_\chi}\right), 
\end{split}
\end{equation}
to the excited $s^P_l=\frac{1}{2}^+$ and $s^P_l=\frac{3}{2}^+$ beauty meson masses, see Eq.~\eqref{rmassesNbD} below. The chiral symmetry breaking scale is $\Lambda_\chi \simeq 1$ GeV and $M$ is $m_\pi$ ($m_K$) for nonstrange (strange) splittings. The magnitude and relative sign of the numerical factors $a$ and $b$ in the above expansion [Eq.~\eqref{dbavggg}] can be chosen to reproduce the observed beauty meson spectrum. 
It is found that for $a$ and $b$ with values less than unity, good results can be obtained; the difference between the observed and predicted masses for the excited $j^P_l=\frac{3}{2}^+$ states is very small, i.e., of order few MeV. 
Theoretical results are not much affected when taking any values for $a$ and $b$ less than unity. Here, we choose $a=0.82$ and $b=-0.18$, which provides more accurate results. In light of the foregoing, one can reexpress the HMChPT masses for the beauty mesons in Eq. ~\eqref{rmassesNb} as
\begin{equation}\label{rmassesNbD}
\begin{split}
&m^r_{B_{(s)}}=\eta_H-\frac{3}{4}\xi^b_H+\alpha_{(s)} L_{H}-\beta_{(s)} F^b_{H}+\Sigma_{B_{(s)}},\\[1ex]
&m^r_{B^*_{(s)}}=\eta_H+\frac{1}{4}\xi^b_H+\alpha_{(s)} L_{H}+\frac{1}{3} \beta_{(s)} F^b_{H}+\Sigma_{B^*_{(s)}},\\[1ex]
&m^r_{B^*_{(s)0}}=\Delta^{(s)}_{SH}+\eta_S-\frac{3}{4}\xi^b_S+\alpha_{(s)} L_{S}-\beta_{(s)} F^b_{S}+\Sigma_{B^*_{(s)0}},\\[1ex]
&m^r_{B^\prime_{(s)1}}=\Delta^{(s)}_{SH}+\eta_S+\frac{1}{4}\xi^b_S+\alpha_{(s)} L_{S}+\frac{1}{3} \beta_{(s)} F^b_{S}+\Sigma_{B^\prime_{(s)1}},\\[1ex]
&m^r_{B_{(s)1}}=\Delta^{(s)}_{TH}+\eta_T-\frac{5}{8}\xi^b_T+\alpha_{(s)} L_{T}-\frac{5}{6} \beta_{(s)} F^b_{T}+\Sigma_{B_{(s)1}},\\[1ex]\
&m^r_{B^*_{(s)2}}=\Delta^{(s)}_{TH}+\eta_T+\frac{3}{8}\xi^b_T+\alpha_{(s)} L_{T}+\frac{1}{2}\beta_{(s)}  F^b_{T}+\Sigma_{B^*_{(s)2}},
\end{split}
\end{equation} 
Both expressions [Eqs. ~\eqref{rmassesNb} and \eqref{rmassesNbD}] will be used in the next section to predict the mass splittings in the beauty sector. The 
absolute masses for the beauty mesons will also be predicted
using Eq. ~\eqref{rmassesNbD}. 

To proceed we need to extract the  values of $(\lambda_{A,1}-\lambda_{H,1})^{(s)}$ and $\delta^{(s)}_{AH}$ using Eqs.~\eqref{dbavg} and \eqref{dbavgg}. From the spectroscopy of the $s^P_l=\frac{1}{2}^-$ and $s^P_l=\frac{3}{2}^+$ charmed and beauty mesons, one finds  $\lambda_{T,1}-\lambda_{H,1}=-0.197~\text{GeV}^2$ ($\delta_{TH}=-54$ MeV) for nonstrange particles and $\lambda^s_{T,1}-\lambda^s_{H,1}=-0.174~\text{GeV}^2$ ($\delta^s_{TH}=-47.7$ MeV) for strange ones. The negative sign shows that the kinetic energy of the heavy quark in the excited $s^P_l=\frac{3}{2}^+$ mesons is larger than that in the $s^P_l=\frac{1}{2}^-$ ground state. From Eq.~\eqref{dbavggg}, the corrections are found to be $\Delta_{TH}=-60$ MeV for nonstrange beauty sector and  $\Delta^s_{TH}=-65$ MeV
for strange one. The extracted values for $\Delta^{(s)}_{TH}$ amount to lowering the masses of the excited $s^P_l=\frac{3}{2}^+$ beauty mesons, see Eq.~\eqref{rmassesNbD}. The nonperturbative parameter $\lambda^{(s)}_{S,1}$ is unknown, and, hence, we cannot extract $\Delta^{(s)}_{SH}$. In Refs. \cite{ms05,HYu}, the value $\lambda^{(s)}_{S,1}-\lambda^{(s)}_{H,1}\approx\lambda^{(s)}_{T,1}-\lambda^{(s)}_{H,1}$ is considered by assuming that the kinetic energy of the heavy quark in the $s^P_l=\frac{1}{2}^+$ states is comparable to that of $s^P_l=\frac{3}{2}^+$ states. To cover all possibilities for the value of the kinetic energy of heavy quark in the $s^P_l=\frac{1}{2}^+$ states and not limit it to values that are comparable to that of the $s^P_l=\frac{3}{2}^+$ states, we consider that the kinetic energy of heavy quark in the $s^P_l=\frac{1}{2}^+$ states lying between those of the $s^P_l=\frac{1}{2}^-$ and $s^P_l=\frac{3}{2}^+$ states and add a large uncertainty that measures our ignorance of $\lambda^{(s)}_{S,1}$. Thus, we take $(\lambda_{S,1}-\lambda_{H,1})^{(s)}=-0.09(6)~\text{GeV}^2$. From Eqs.~\eqref{dbavgg} and \eqref{dbavggg}, the beauty nonstrange (strange) $s^P_l=\frac{1}{2}^+$ masses in Eq.~\eqref{rmassesNbD} are lowered by $\Delta_{SH}=-27(18)$ MeV [$\Delta^{s}_{SH}=-33(22)$ MeV].
At our level of precision, the uncertainties due to experimental masses and higher order $O(\Lambda_{\text{QCD}}/m_c-\Lambda_{\text{QCD}}/m_b)$ corrections are negligible.

\section{results and discussion}
Before presenting our results, let us first illustrate the fitting and predicting methods. Here, we follow the approach employed in \cite{Alhakami} to fit $\eta$, $\xi$, $L$, and $F$ parameters and predict the beauty meson masses. The fitting method is essentially based on using physical masses of charmed mesons to evaluate the chiral loop functions, which ensures that the imaginary
parts of self-energies (loop functions) are correctly related to the observed decay widths. This makes fit linear and hence helps to extract unique values for the $\eta$, $\xi$, $L$, and $F$ parameters.
Because of computing loop integrals using physical masses,
the obtained values for the parameters will contain contributions beyond $\mathcal{O}(Q^3)$. The generated higher order $\mu$-dependent terms cannot properly be renormalized using $\mu$-dependence counterterms of the theory. Therefore, a theoretical error coming from such higher order terms should be estimated. As in \cite{Alhakami}, the $\beta$ functions of the $\eta$, $\xi$, $L$, and $F$ parameters will be defined to estimate the contributions from the generated higher-order terms.

In the fit, we use twelve masses of strange and nonstrange charmed mesons \cite{pdg12}, see Fig. \ref{Ddoublets}. We work in the isospin limit. We only average the masses of the well-determined charmed nonstrange mesons. The $\frac{1}{2}^+$ charmed nonstrange mesons, however, are poorly determined. So, we instead use the masses of the excited $D_0^{*0}$, 2300(19) MeV \cite{pdg12}, and $D^{0\prime}_1$, 2427(26)(25) MeV \cite{19}, mesons. For Goldstone particles, $m_{\pi}=140$ MeV, $m_K = 495$ MeV, and $m_\eta = 547$ MeV are used. 
Our results smoothly change with the normalization scale $\mu$;
consequently, performing calculations at any other values of the normalization scale will not make much difference. In our numerical calculations, we set the normalization scale to the average of pion and kaon masses, $\mu = 317$ MeV, as in \cite{Alhakami}.
The numerical values for the couplings can be extracted using available data on strong decays of charmed mesons. The coupling constant $g$ at tree level can be extracted using the measured width of $D^{*\pm}$ \cite{pdg12}; this gives $g=0.5672(80)$. For the $h$ coupling, we use $h=0.514(17)$, which is extracted from the width of $D^{*\pm}_0$ \cite{cheng}. The $g^\prime$, $h^\prime$, and $g^{\prime\prime}$ couplings are unknown experimentally.  We, therefore, use lattice QCD result for $g^\prime=-0.122(8)(6)$ \cite{glattice} and restrict $h^\prime$ and $g^{\prime\prime}$ to lie between 0 and 1; so, one can study the variation of the calculated masses with the $h^\prime$ and $g^{\prime\prime}$ couplings. By confronting our resulting masses against experiments, these unmeasured couplings will be constrained to lie in a narrow range making our theory much reliable. 

As our calculations are performed at different $h^\prime$ and $g^{\prime\prime}$, we will only show the fitting method considering $h^\prime=0.30$ and $g^{\prime\prime}=0.03$. In Eq.~\eqref{rmassesN}, $m_A^r$ represents the residual mass of the charmed meson $A$, which is taken to be the difference between the experimental mass and an arbitrarily chosen reference mass of $O(m_c)$ \cite{ms05}. Here we choose $m_{D^*}$ as a reference, which yields the following central values for charmed meson residual masses
\cite{pdg12,19}
\begin{equation}\label{rm0}
\begin{split}
m^r_D&=-141.32, ~~m^r_{D^*}=0,~~~~~~~~~m^r_{D_s}=-40.22,\\
m^r_{D^*_s}&= 103.65,~~~~m^r_{D_0^*}=291.45,~~~m^r_{D^\prime_1}= 418.45,\\
m^r_{D^*_{s0}}&=309.25,~~~ m^r_{D^\prime_{s1}}=450.95,~~~m^r_{D_1}=413.45,\\
m^r_{D_{s1}}&= 526.56,~~~~m^r_{D^*_2}=454.5,~~~~ m^r_{D^*_{s2}}=560.55
\end{split}
\end{equation}
in MeV units. By fitting the mass expressions in Eq.~\eqref{rmassesN} to the corresponding empirical masses  in Eq.~\eqref{rm0}, one obtains
\begin{equation}\label{rm1}
\begin{split}
\eta_H&=104(7),~~~~~~\xi_H=149(5),\\
L_H&=212(14),~~~~F_H=-44(11),\\
\eta_S&=385(19),~~~~~\xi_S=138(27),\\
L_S&=20(31),~~~~~F_S=12(42),\\
\eta_T&=490(1),~~~~~~\xi_T=41(1),\\
L_T&=119(1),~~~~~F_T=-4(2),
\end{split}
\end{equation}
which are given in MeV units. The fit results are obtained by computing the chiral loop functions in Eq.~\eqref{rmassesN} using physical masses and couplings, as mentioned above. The errors use to get this fit are the experimental errors on masses and couplings, the lattice QCD error on $g^\prime$, and the theoretical error that we have estimated from the $\beta$  functions of the parameters. The uncertainty in the nonstrange $D_0^{*0}$ and $D^{0\prime}_1$ masses gives rise to the large uncertainties seen in $\eta_S,~\xi_S,~L_S$, and $F_S$ parameters.

The charm sector results on HMChPT parameters, Eq.~\eqref{rm1}, can be used in Eq.~\eqref{rmassesNb} [Eq.~\eqref{rmassesNbD}]
to predict the mass splittings (the mass splittings and absolute masses) in the beauty sector. For an illustration, we will use Eq.~\eqref{rmassesNbD} to extract the beauty meson masses. 
Following \cite{Alhakami}, one can choose the ground state, $B$, as a reference mass to define eleven independent mass splittings, $\Delta m_A=m_A-m_B$, where $A\in \{B^*,B_s,B^*_s,B^*_0,B^\prime_1,B^*_{s0},B^\prime_{s1},
B_1,B^*_2,B_{s1},B^*_{s2}\}$. As the self-energies represent nonlinear functions of the beauty meson mass differences, these independent splittings form nonlinear equations. An iterative method is utilized to solve them starting from the tree-level masses. Adding the mean value of the observed mass $m_B$, which are chosen as a reference mass in our calculations, to the predicted mass splittings, yields
\begin{equation}
\begin{split}\label{predict}
m_{B^*}&=5325(9),~~~~~~~~~~~~m_{B_s}=5369(13),\\
m_{B_s^*}&=5415(15),~~~~~~~~~~m_{B_0^*}=5681(25)(18),\\
m_{B_1^\prime}&=5719(26)(18),~~~~m_{B_{s0}^*}=5711(32)(22),\\
m_{B_{s1}^\prime}&=5756(31)(22),~~~~~m_{B_1}=5726(13),\\
m_{B_2^*}&=5739(14),~~~~~~~~~~m_{B_{s1}}=5830(14),\\
m_{B_{s2}^*}&=5840(14)
\end{split}
\end{equation}
in MeV units. The  errors (first for the $s^P_l=\frac{1}{2}^+$ masses) come from the uncertainties in the charmed masses, coupling constants, lattice QCD computation on $g^\prime$, rescaling factor, and estimated theoretical error. The first error on the predicted masses of the excited $s^P_l=\frac{1}{2}^+$ beauty mesons is dominated by the uncertainty in $D_0^{*0}$ and $D^{0\prime}_1$ masses and the second is the uncertainty in $\lambda^{(s)}_{S,1}$.

To investigate the influence of virtual loop effects on the predicted beauty meson spectrum, our calculations  are performed considering different input values for $h^\prime$ and $g^{\prime\prime}$. The spectrum of the $s^P_l=\frac{1}{2}^-$ and $s^P_l=\frac{1}{2}^+$ beauty mesons depend only on the $h^\prime$ coupling. Both $h^\prime$ and $g^{\prime\prime}$ couplings, however, affect the excited $s^P_l=\frac{3}{2}^+$ beauty meson masses. The results can be justified by confronting them with the experiments. 
\begin{figure}[h!]
\includegraphics[width = 0.7\textwidth]{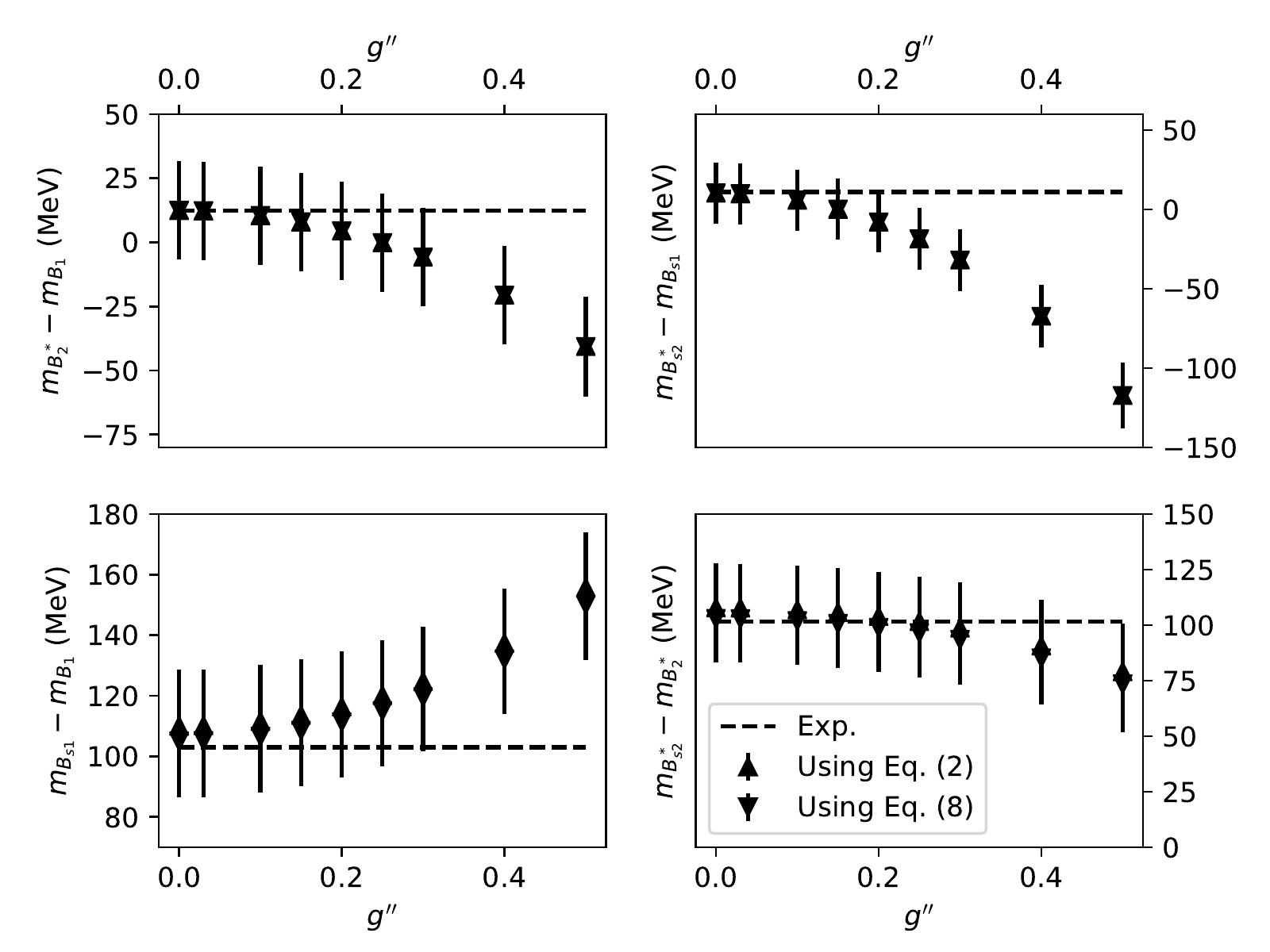}
\caption{The predicted hyperfine splittings (upper panel) and $SU(3)$ flavor splittings (lower panel) in the excited $s^P_l=\frac{3}{2}^+$ doublets using HMChPT masses in Eqs.~\eqref{rmassesNb} and \eqref{rmassesNbD} plotted against $g^{\prime\prime}$. The associated errors include the experimental errors on the charmed masses
and coupling constants, the error from the lattice QCD calculations of $g^\prime$, the error from the input parameter $\frac{m_c}{m_b}$, and the estimated theoretical error. Different symbols are given to the experiment and two theoretical predictions using HMChPT masses in Eqs.~\eqref{rmassesNb} and \eqref{rmassesNbD} according to the key in the ($m_{B^*_{s2}}-m_{B^*_2}$) plot.}
\label{Tsector}
\end{figure}

We first look at the dependence of the higher excited $s^P_l=\frac{3}{2}^+$ states on $g^{\prime\prime}$. To neglect loops effect from the virtual excited $s^P_l=\frac{1}{2}^+$ states, the $h^\prime$ coupling is set to zero. As shown in Fig.~\ref{Tsector}, the self-energy effects within $s^P_l=\frac{3}{2}^+$ states, which are parametrized by $g^{\prime\prime}$, have a strong impact on the predicted masses. The calculated  masses of $B_1$ and $B_{s1}$ ($B^*_2$ and $B^*_{s2}$) mesons significantly increase (decrease) as $g^{\prime\prime}$ increases. The observations and predictions are in a good agreement for small values of the coupling, $g^{\prime\prime}< 0.10$. At $g^{\prime\prime}\simeq 0.15$ ($g^{\prime\prime}\simeq 0.25$), the excited $B_{s1}$ and $B^*_{s2}$ ($B_1$ and $B^*_2$) states become degenerate; which in turn implies that the predictions are unreliable in the limit $g^{\prime\prime}>0.10$. Accordingly, $g^{\prime\prime}$ should be constrained to lie between $0$ and $0.10$. Hyperfine splittings are independent of $O(\Lambda_{\text{QCD}}/m_c-\Lambda_{\text{QCD}}/m_b)$ corrections, and this is confirmed in our theoretical results. However, there is no exact cancellation between the $O(\Lambda_{\text{QCD}}/m_c-\Lambda_{\text{QCD}}/m_b)$ corrections to nonstrange and strange heavy meson masses. Such corrections lower the predicted $SU(3)$ splittings [compare results from Eqs.~\eqref{rmassesNb} and \eqref{rmassesNbD}] by nearly 5 MeV with central values close to the observed ones for $g^{\prime\prime}< 0.10$. In the following, we will use $g^{\prime\prime}=0.03$.

\begin{figure}[h!]
\subfloat[ ]{\includegraphics[width = 0.7\textwidth]{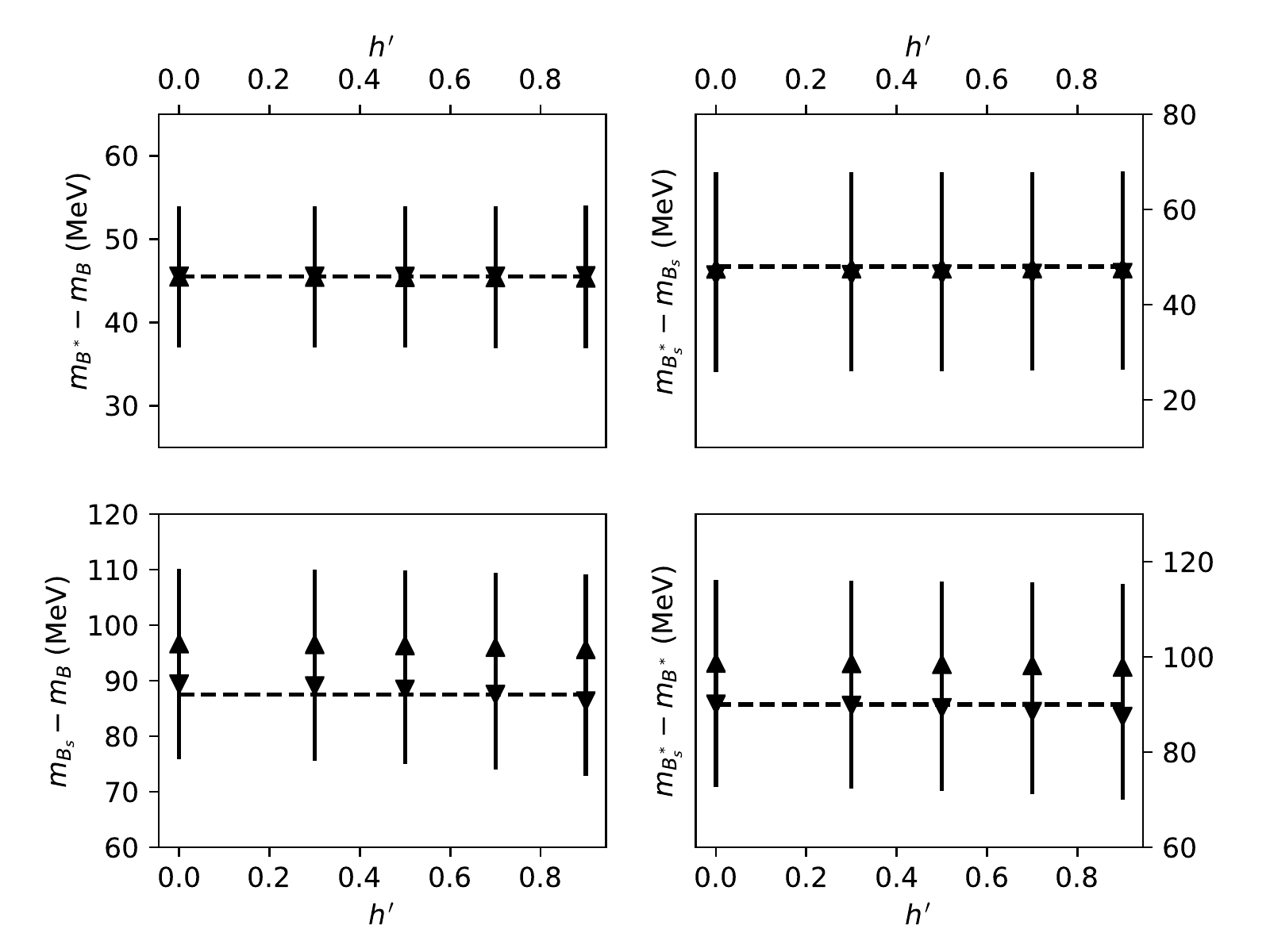}} \\
\subfloat[ ]{\includegraphics[width = 0.7\textwidth]{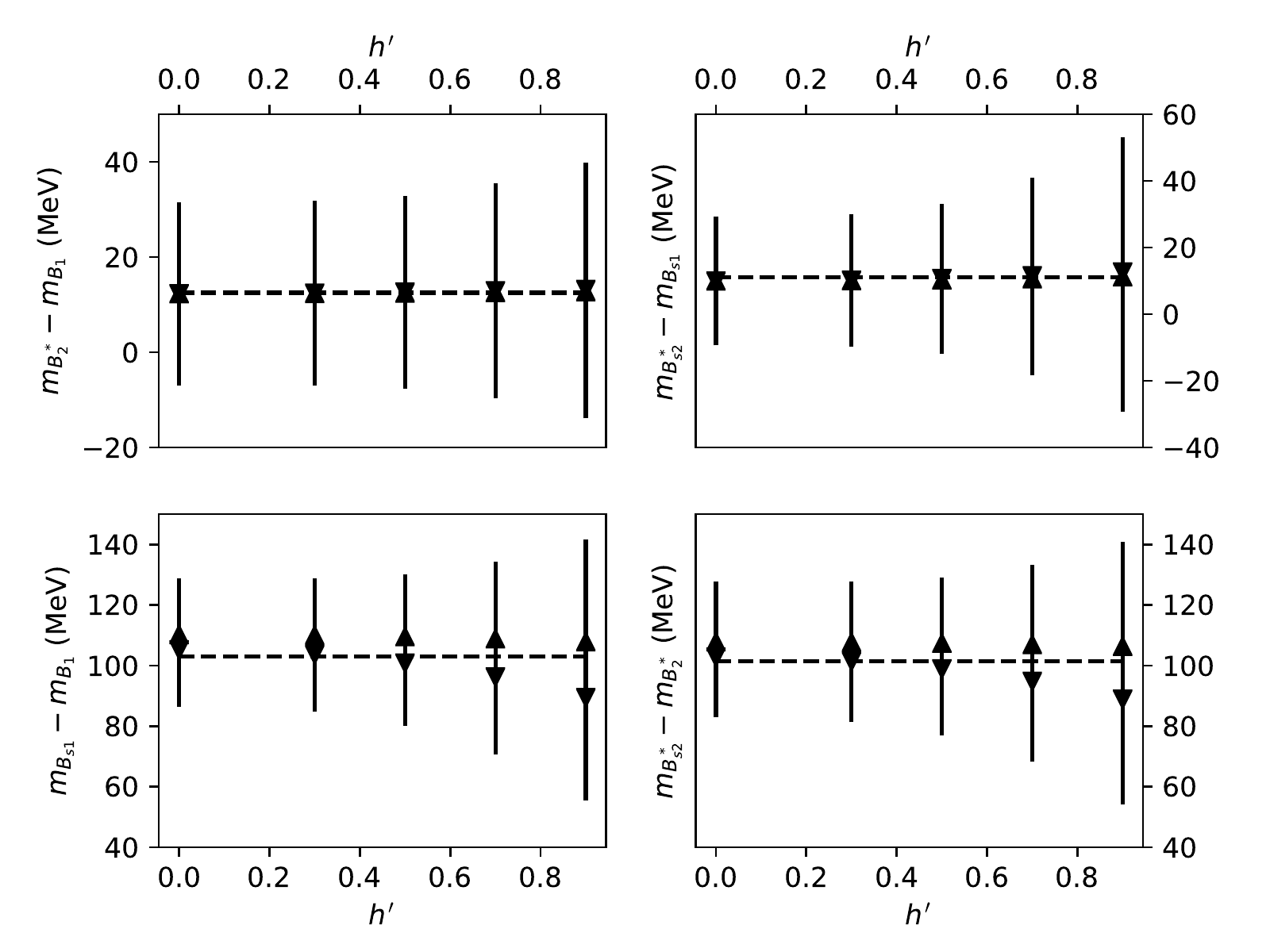}}
\caption{The predicted hyperfine splittings (upper panel) and $SU(3)$ flavor splittings (lower panel) in the (a) $s^P_l=\frac{1}{2}^-$ and (b) $s^P_l=\frac{3}{2}^+$ heavy quark doublets plotted against $h^\prime$.
The notation is the same as in Fig.\ref{Tsector}.}
\label{HTp}
\end{figure}

\begin{figure}[h!]
\includegraphics[width = 0.7\textwidth]{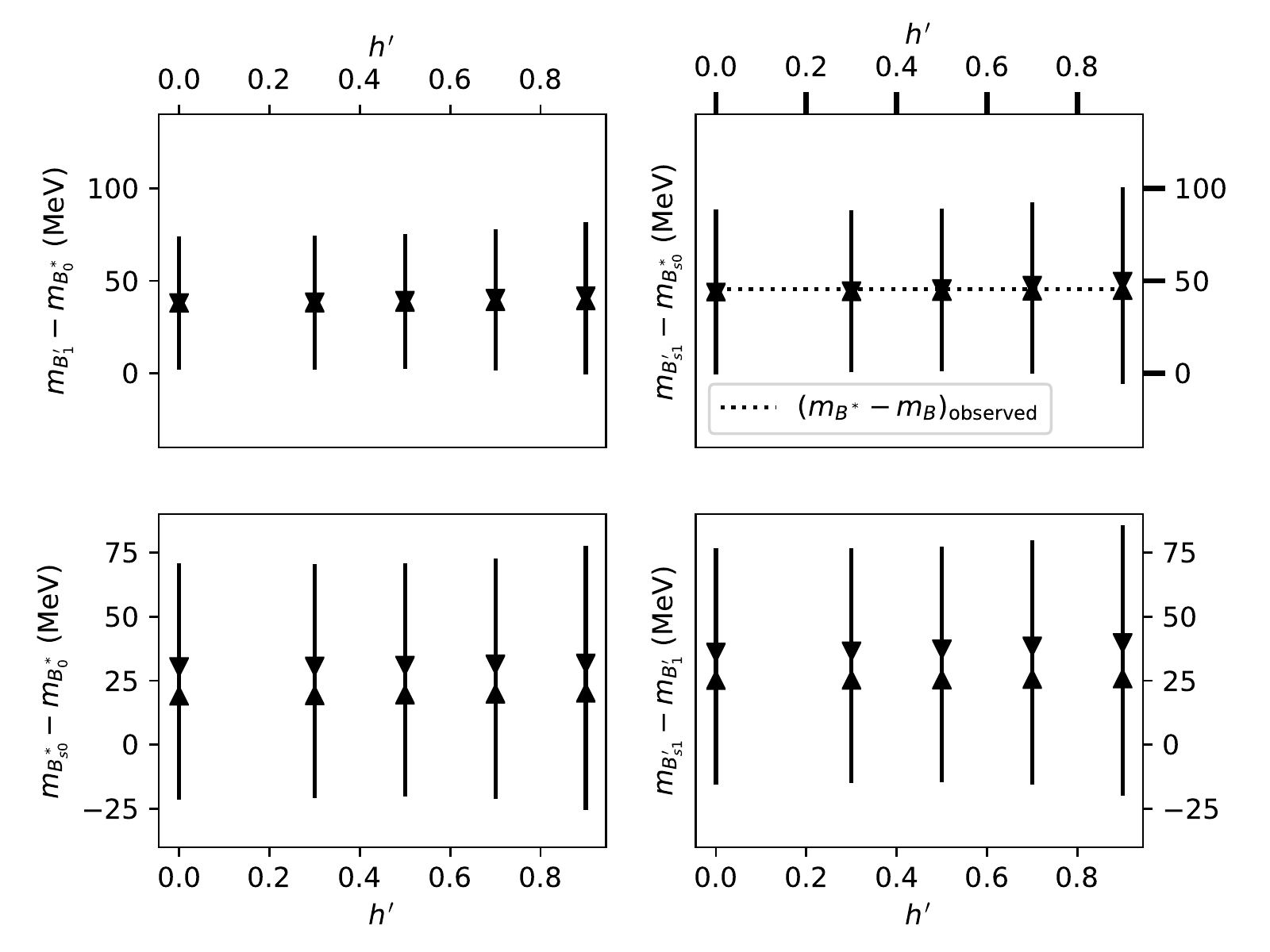}
\caption{The predicted hyperfine splittings (upper panel) and $SU(3)$ flavor splittings (lower panel) in the excited $s^P_l=\frac{1}{2}^+$ heavy quark doublets plotted against $h^\prime$. The notation is the same as in Fig.\ref{Tsector}. These beauty mesons have not yet been discovered. Our predictions (upper right panel) confirm the mass relation $m_{B^\prime_{s1}}-m_{B^*_{s0}}\approx m_{B^*}-m_B \approx m_{B^*_s}-m_{B_s}$, which has been observed in the charm sector and expected in the beauty sector by heavy quark symmetry; for details please
refer to the text.}
\label{Sp}
\end{figure}

The variation of the predicted beauty meson splittings with $h^\prime$ is presented in Figs.~\ref{HTp} and \ref{Sp} and compared with the available experimental data. In Fig.~\ref{HTp}, the predicted hyperfine splittings in the $s^P_l=\frac{1}{2}^-$ and $s^P_l=\frac{3}{2}^+$ doublets, which exhibit weak dependence on $h^\prime$, are in excellent agreement with the observed values.
Evidently, the $O(\Lambda_{\text{QCD}}/m_c-\Lambda_{\text{QCD}}/m_b)$ corrections will, in general, lower the $SU(3)$ flavor splittings in 
the $s^P_l=\frac{1}{2}^-$ and $s^P_l=\frac{3}{2}^+$ doublets
by nearly 10 and 5 MeV, respectively. In the lower panel of Fig.~\ref{HTp}(a), the results for the flavor splittings in the $s^P_l=\frac{1}{2}^-$ doublet obtained using Eqs.~\eqref{rmassesNb} and \eqref{rmassesNbD} show weak dependence on $h^\prime$. The extracted values from applying Eq.~\eqref{rmassesNbD}, which takes into account $O(\Lambda_{\text{QCD}}/m_c-\Lambda_{\text{QCD}}/m_b)$ corrections, are in excellent agreement with the reported values. However, for those obtained using Eq.~\eqref{rmassesNb}, the agreement are within 1$\sigma$ standard deviation. 
For the $SU(3)$ flavor splittings in the excited $s^P_l=\frac{3}{2}^+$  doublet, the theoretical results are in good agreement with data for $h^\prime\leqslant0.5$, as shown in the lower panel of Fig.~\ref{HTp}(b).  For $h^\prime>0.5$, the results (more precisely the central values) extracted using Eq.~\eqref{rmassesNbD} start to deviate from the observations. Therefore, the coupling $h^\prime$ should be restricted to values smaller than $0.5$. The Fig.~\ref{HTp}(b) also shows how theoretical error, which is estimated from the $\beta$ functions of the parameters, varies with $h^\prime$.

The $s^P_l=\frac{1}{2}^+$ beauty mesons have not yet been observed. Our theoretical calculations of hyperfine and flavor splittings for these missing states are shown in Fig.~\ref{Sp}. The empirical relation $m_{D^\prime_{s1}}-m_{D^*_{s0}}\approx m_{D^*}-m_D \approx m_{D^*_s}-m_{D_s}$ in the charm sector is expected in the beauty sector by heavy quark symmetry \cite{pdg12,cheng}. 
Our predictions confirm that $m_{B^\prime_{s1}}-m_{B^*_{s0}}\approx m_{B^*}-m_B \approx m_{B^*_s}-m_{B_s}$, as shown in the upper right panel of Fig.~\ref{Sp}.
Our calculations show that there is an accidental cancellation
between counterterms and $SU(3)$ breaking loop corrections
in the predicted flavor splittings.
The predicted $SU(3)$ splittings 
$m_{B^*_{s0}}-m_{B^*_0}\simeq 19$ MeV and $m_{B^\prime_{s1}}-m_{B^\prime_1}\simeq 25$ MeV ($m_{B^*_{s0}}-m_{B^*_0}\simeq 30$ MeV and $m_{B^\prime_{s1}}-m_{B^\prime_1}\simeq 36$ MeV) are obtained using Eq.~\eqref{rmassesNb} [Eq. \eqref{rmassesNbD}]. These results are far below theoretical expectations like those in the charm sector; i.e., $m_{D^*_{s0}}-m_{D^{*0}_0}\approx 18$ MeV and $m_{D^\prime_{s1}}-m_{D^{0\prime}_1}\approx 33$ MeV \cite{pdg12,19}. Thus, our results are consistent with the expectations of heavy quark spin-flavor symmetry.

We have shown the predicted hyperfine and flavor splittings from applying HMChPT masses in Eq. \eqref{rmassesNbD}, which incorporates $O(\Lambda_{\text{QCD}}/m_c-\Lambda_{\text{QCD}}/m_b)$ corrections, in Figs.~
\ref{Tsector}--\ref{Sp}. 
The variation of the predicted beauty meson
masses with $h^\prime$ is presented in Table \ref{tab:2} and compared with the experimental data. The masses of the $s^P_l=\frac{1}{2}^-$ and nonstrange $s^P_l=\frac{3}{2}^+$ ($s^P_l=\frac{1}{2}^+$ and strange $s^P_l=\frac{3}{2}^+$) states show a weak (strong) dependence on the $h^\prime$ coupling. For $B^*$, $B_s$, $B^*_s$, $B_1$, and $B^*_2$ states, the difference between the mean values of the observed and predicted masses is very small, i.e., of order few MeV. However, this is not the case for the $s^P_l=\frac{1}{2}^+$ and strange $s^P_l=\frac{3}{2}^+$ beauty mesons, where the virtual loop effects lower their masses  by an amount of $\sim O(15)$ MeV. 

\begin{table*}[ht!]
\def\arraystretch{2.5}
\begin{center}
\caption{Comparison of the experimental data \cite{pdg12} and our theoretical results
using HMChPT masses in Eq.~\eqref{rmassesNbD}. We take the isospin average of $B^-_1$ and $B^0_1$ ($B^{*-}_2$ and $B^{*0}_2$) to obtain the mass of nonstrange excited state $B_1$ ($B^*_2$). In our calculations, we fix $g^{\prime\prime}=0.03$ and use different values for the $h^\prime$ coupling.
Masses are in units of MeV.}
\begin{tabular}{lllllllllc}
\hline\hline
\multicolumn{10}{c}{Theoretical predictions at different $h^\prime$} \\
\cline{4-7}
$j^P_l$&$J^P$&Meson&$0.00$&$0.30$&$0.50$&$0.90$&Experiment&   &\\ \hline
$\frac{1}{2}^-$&$1^-$&$B^*$&5325(9)&5325(9)&5325(9)&5325(9)&5324.70(21)&   &  \\  
$\frac{1}{2}^-$&$0^-$&$B_s$&5369(13)&5369(13)&5368(13)&5366(13)&5366.88(14)&   &\\  
$\frac{1}{2}^-$&$1^-$&$B_s^*$&5415(15)&5415(15)&5414(15)&5413(15)&5415.4(2.3)&  &\\  
$\frac{1}{2}^+$&$0^+$&$B^*_0$&5683(25)(18)&5681(25)(18)&5678(25)(18)&5667(28)(18)&unseen&   &  \\  
$\frac{1}{2}^+$&$1^+$&$B_1^\prime$&5721(26)(18)&5719(26)(18)&5717(26)(18)&5709(28)(18)&unseen&  &  \\  
$\frac{1}{2}^+$&$0^+$&$B^*_{s0}$&5713(32)(22)&5711(32)(22)&5709(32)(22)&5699(36)(22)&unseen&  &  \\  
$\frac{1}{2}^+$&$1^+$&$B_{s1}^\prime$&5757(31)(22)&5756(31)(22)&5754(31)(22)&5749(36)(22)&unseen&  &  \\  
$\frac{3}{2}^+$&$1^+$&$B_1$&5727(13)&5726(13)&5725(14)&5724(19)&5726(2)&   &\\  
$\frac{3}{2}^+$&$2^+$&$B_2^*$&5739(14)& 5739(14)&5738(15)&5737(19)&5738.35(49)&   &\\  
$\frac{3}{2}^+$&$1^+$&$B_{s1}$&5832(13)&5830(14)&5826(15)&5813(28)&5828.7(2)&    &\\  
$\frac{3}{2}^+$&$2^+$&$B_{s2}^*$&5842(14)&5840(14)&5837(16)&5826(29)&5839.86(12)&   &\\  
\hline\hline
\end{tabular}
\label{tab:2}
\end{center}
\end{table*}
By analyzing the predicted and observed spectrum in Table \ref{tab:2}, it is clear that  the $s^P_l=\frac{1}{2}^-$ and $s^P_l=\frac{3}{2}^+$ states are well reproduced, which without a doubt reflects the power of HMChPT. Clearly, the best values for the masses of the strange excited $s^P_l=\frac{3}{2}^+$ states are those extracted considering $h^\prime\leqslant0.50$. It is worth remarking that the predicted excited $s^P_l=\frac{1}{2}^+$ beauty nonstrange (strange) masses are well above (below) the threshold for decays to ground state $B$ mesons and pions (kaons), and therefore these mesons are expected to be very broad (narrow) like $s^P_l=\frac{1}{2}^+$ charmed nonstrange (strange) mesons.  Our predictions for the not yet discovered $B^*_{s0}$ and $B^\prime_{s1}$ states are remarkably close to the lattice QCD results in \cite{B}: $m_{B^*_{s0}}=5711(13)(19)$ MeV and $m_{B^\prime_{s1}}=5750(17)(19)$ MeV.

It should be noted that the approach employed in the current paper and \cite{Alhakami} to predict the beauty meson spectrum within HMChPT framework is different from the one utilized in \cite{HYu,cheng,col}. In this work and in \cite{Alhakami}, a third-order one-loop mass expansion for the heavy-light mesons, which takes into account the effects to first order in light quark mass, $m_q$, and to first order in $1/m_Q$, and $m_q/m_Q$ terms, is considered. The unknown parameters in such mass expansion, i.e., Eq. \eqref{rmassesN}, are fixed using charm spectrum. Then, the fitted hyperfine parameters, which are functions of $\mathcal{O}(1/m_c)$, are scaled by $m_c/m_b$ to define the mass expansion, i.e., Eqs. \eqref{rmassesNb} and  \eqref{rmassesNbD}, for the corresponding beauty meson states. A self-consistent approach is then used to extract the beauty meson masses. In \cite{HYu,cheng,col}, on the other hand, the treatment is based on the heavy quark symmetry argument. In these studies, the one-loop corrections and chiral symmetry breaking terms are neglected. In the rest of this section, we will briefly review and discuss the approach employed in these studies. For this, we will use the notation of \cite{col}
and restrict the discussion to the missing excited $s^P_l=\frac{1}{2}^+$ states. In the strict HQ limit \cite{col},
\begin{equation}
\begin{split}\label{col}
\Delta^{(c)}_S &=\Delta^{(b)}_S,\\
\lambda^{(c)}_S &=\lambda^{(b)}_S,\\
\end{split}
\end{equation}
where $\Delta_S$ measures the spin-averaged mass splittings between the excited $S$ doublet and ground state $H$ doublet (this corresponds to $\delta_S-\delta_H$ in \cite{ms05,Alhakami,Alhakami20})
and $\lambda_S$ measures the mass splitting between spin partners of $S$ doublet. The quantity $\lambda_S$ represents the HQET  nonperturbative parameter $\lambda_{S,2}$, see Eq.~\eqref{hqet}. The HMChPT hyperfine parameter, which is denoted by $\Delta_S$ in \cite{ms05,Alhakami,Alhakami20}, is related to this HQET parameter by $\Delta_S=\lambda_S/m_c$ for the case of the charmed mesons. Note that the mass expansion in \cite{ms05,Alhakami,Alhakami20} is defined up to to first order in $1/m_c$, which in turn implies that $\lambda^{(c)}_S\equiv\lambda^{(b)}_S$. Therefore, it is sufficient to rescale the hyperfine parameters ($\xi$ and $F$) in the charmed meson mass expansion [Eq.~\eqref{rmassesN}] by the mass ratio $m_c/m_b$ to define the HMChPT mass expansion for the analog beauty meson states, i.e., Eqs. \eqref{rmassesNb} and  \eqref{rmassesNbD}. The leading QCD corrections and $1/m_Q$ effect \cite{lamdc} to $\lambda^{(c)}_X\equiv\lambda^{(b)}_X$
are beyond the order, i.e., $\mathcal{O}(Q^3)$, to which we are working. Such corrections, which have been considered
in \cite{cheng,HYu}, are found to shift the masses by a few MeV; see the text below Eq.~\eqref{2014}.

The lhs of the relations in Eq.~\eqref{col} is experimentally determined. The two relations are then used to predict states in the corresponding beauty doublets. For the missing excited $s^P_l=\frac{1}{2}^+$ states, the authors of \cite{col} obtained
\begin{equation}
\begin{split}\label{12}
m_{B^*_0}&=5708.2(22.5),~~~m_{B^\prime_1}=5753.3(31.1),\\
m_{B^*_{s0}}&=5706.6(1.2),~~~m_{B^\prime_{s1}}=5765.6(1.2) 
\end{split}
\end{equation}
in MeV units. The leading QCD corrections and $1/m_Q$ effect to the relations in Eq.~\eqref{col} are taken into account
in \cite{cheng,HYu}. The leading $1/m_Q$ correction to $\Delta^{(c)}_S=\Delta^{(b)}_S$ is estimated using HQET as done above. It is given by 
\begin{equation}\label{kk}
 \Delta^{(b)}_S=\Delta^{(c)}_S+\delta_{SH},
\end{equation}
where $\delta_{SH}\sim \mathcal{O}(-35)$ MeV is obtained using $4.65$ and $1.275$ GeV for the beauty and charm quark masses, respectively, and by assuming that the kinetic energy of the heavy quark in the $s^P_l=\frac{1}{2}^+$ states is comparable to that of $s^P_l=\frac{3}{2}^+$ states. 
The next-to-leading corrections to the relation
$\lambda^{(c)}_S=\lambda^{(b)}_S$ has not yet been calculated.  
The authors of \cite{HYu} have generalized the calculations of Amoros, Beneke, and Neubert \cite{lamdc}
for the negative-parity mesons by considering the leading QCD corrections to the relation $\lambda^{(c)}_S=\lambda^{(b)}_S$, i.e., \begin{equation}\label{hl}
\lambda^{(b)}_S=\lambda^{(c)}_S\left(\frac{\alpha_s{(m_b)}}{\alpha_s{(m_c)}}\right)^{9/25}. 
\end{equation}
With aforementioned improvements [Eqs.~\eqref{kk} and \eqref{hl}], the authors of \cite{HYu} obtained (in MeV)
\begin{equation}
\begin{split}\label{14}
m_{B^*_0}&=5715(22)+\delta_{SH},~~~m_{B^\prime_1}=5752(31)+\delta_{SH},\\
m_{B^*_{s0}}&=5715(1)+\delta_{SH},~~~m_{B^\prime_{s1}}=5763(1)+\delta_{SH}, 
\end{split}
\end{equation}
where the $\delta_{SH}$ correction amounts to lowering the masses of the excited $s^P_l=\frac{1}{2}^+$ beauty states.

The predicted masses in \cite{col,cheng,HYu}, which are given in Eqs.~\eqref{12} and \eqref{14}, are obtained using $m_{D^{*0}_0}=2318(29)$ MeV for the nonstrange scalar charmed meson, which is very close to its strange partner, $D^{*+}_{s0}$. Our theoretical predictions in Figs.~\ref{Tsector}--\ref{Sp} and Table \ref{tab:2} are obtained 
using the 2020 PDG \cite{pdg12}, where the updated value for the mass of the scalar $D^{*0}_0$ state is $2300(19)$ MeV, which is smaller than $D^{*+}_{s0}$ by nearly 18 MeV. To compare the approach employed in this work with the one used in \cite{col,cheng,HYu}, the results in 
Eqs.~\eqref{12} and \eqref{14} must be updated
using the 2020 PDG \cite{pdg12}. This leads to the following predicted masses (in MeV)
\begin{equation}
\begin{split}\label{2012}
m_{B^*_0}&=5696(19),~~~~~~~m_{B^\prime_1}=5749(31),\\
m_{B^*_{s0}}&=5706.9(1.8),~~~m_{B^\prime_{s1}}=5765.9(1.8), 
\end{split}
\end{equation}
in the HQ limit [Eq.~\eqref{col}] and 
\begin{equation}
\begin{split}\label{2014}
m_{B^*_0}&=5702(20)+\delta_{SH},~~~m_{B^\prime_1}=5746(30)+\delta_{SH},\\
m_{B^*_{s0}}&=5714.0(1.8)+\delta_{SH},\\
m_{B^\prime_{s1}}&=5763.5(1.8)+\delta_{SH}, 
\end{split}
\end{equation}
when considering the $1/m_Q$ [Eq.~\eqref{kk}] and
leading QCD [Eq.~\eqref{hl}] corrections. By comparing Eqs.~\eqref{2012} and \eqref{2014}, it is obvious that including the leading QCD corrections enhances (reduces) the masses of the scalar (axial-vector) beauty mesons by nearly 7 MeV (2 MeV), as already noted in \cite{HYu,cheng}. The predicted $B^*_{s0}$ and $B^\prime_{s1}$ states in Eq.~\eqref{2014} are close to our predictions in Table \ref{tab:2}. However, the masses for $B^*_0$ and $B^\prime_1$ states in Eq.~\eqref{2014} are higher than the predicted masses in Table \ref{tab:2} by order 20 and 30 MeV, respectively. 
\section{CONCLUSIONS}
The spectroscopy of the ground-state ($s_l^P=\frac{1}{2}^-$) and lowest-excited ($s_l^P=\frac{1}{2}^+$ and $s_l^P=\frac{3}{2}^+$) beauty mesons were analyzed within HMChPT framework. The mass expressions used in our study include all leading contributions from one-loop corrections and those due to chiral and  heavy quark symmetry breakings. The charmed spectrum was used to fix the unknown parameters that appear in the mass formulas. Then, we used charm results to make predictions for the analog beauty meson spectrum. 

Our calculations were performed at different values for the experimentally and theoretically unknown ($h^\prime$ and $g^{\prime\prime}$) couplings, which helped to examine the influence of virtual loops effect on the calculated masses. It was found that the data is more consistent with the predicted
$s_l^P=\frac{3}{2}^+$ masses when $g^{\prime\prime}<0.10$. For self-energy corrections parametrized by $h^\prime$, the calculated masses for the $s^P_l=\frac{1}{2}^-$ and nonstrange $s^P_l=\frac{3}{2}^+$ ($s^P_l=\frac{1}{2}^+$ and strange $s^P_l=\frac{3}{2}^+$) beauty mesons were found to have a weak (strong) dependence on $h^\prime$. The $s^P_l=\frac{1}{2}^+$ and strange $s^P_l=\frac{3}{2}^+$ beauty mesons were pushed down by nearly $O(15)$ MeV. The resulting masses for the $s_l^P=\frac{1}{2}^-$ and $s_l^P=\frac{3}{2}^+$
beauty mesons are consistent with the observed values. 
However, the $s^P_l=\frac{1}{2}^+$ beauty mesons have not yet been discovered; so, our findings could provide useful information for experimentalists investigating such states.  

\section{Acknowledgments}
The author extends his appreciation to the Deanship of Scientific Research at King Saud University for funding this work through Research  Group No. RG-1441-537.

\end{document}